\documentclass[aps,pre,showpacs,twocolumn,groupedaddress,floatfix]{revtex4}
\usepackage{amsmath}
\usepackage{amssymb}
\usepackage{epsfig}
\usepackage{amscd}
\usepackage{color}
\usepackage{comment}
\usepackage{ulem}
\usepackage{bm}

\newcommand{\BEC}{\text{BEC}}
\newcommand{\cl}{\text{cl}}
\newcommand{\qc}{\text{qc}}
\newcommand{\fs}{\text{fs}} 
\newcommand{\TBECZERO}{T^{2\text{d}}_{\text{BEC}}}

\newcommand{\KT}{\text{KT}}

\newcommand{\elabel}[1]{\label{e:#1}}
\newcommand{\slabel}[1]{\label{s:#1}}
\newcommand{\eq}[1]{Eq.~(\ref{e:#1})}

\newcommand{\mutilde}{\tilde{\mu}}
\newcommand{\rtilde}{\tilde{r}}
\newcommand{\ntilde}{\tilde{n}}
\newcommand{\ztilde}{\tilde{z}}
\newcommand{\epstilde}{\tilde{\epsilon}}
\newcommand{\omegaztilde}{\tilde{\omega}_z}
\newcommand{\gtilde}{\tilde{g}}
\newcommand{\phitilde}{\tilde{\phi}}

\newcommand{\Psitilde}{\tilde{\Psi}}

\newcommand{\eqtwo}[2]{Eqs~(\ref{e:#1}) and~(\ref{e:#2})}
\newcommand{\fig}[1]{Fig.~\ref{f:#1}}
\newcommand{\figg}[1]{Figure~\ref{f:#1}}
 
\newcommand{\sect}[1]{Section~\ref{s:#1}}

\newcommand{\quot}[1]{``#1''}

\newcommand{\SET}[1]{\{#1\}}

\newcommand{\VEC}[1]{\mathbf{#1}}
\newcommand{\avec}{\VEC{a}}
\newcommand{\Avec}{\VEC{A}}

\newcommand{\kvec}{\VEC{k}}

\newcommand{\rvec}{\VEC{r}}
\newcommand{\rvectilde}{\tilde{\VEC{r}}}

\newcommand{\rvecD}{\vec{r}\,}
\newcommand{\rvecDP}{\vec{r}\,'}

\newcommand{\expa}[1]{\mathrm{e}^{#1}}   
\newcommand{\expb}[1]{\exp \glb #1 \grb} 
\newcommand{\expc}[1]{\exp \glc #1 \grc} 
\newcommand{\logb}[2][]{\log^{#1} \glb #2 \grb}  
\newcommand{\logc}[2][]{\log^{#1} \glc #2 \grc}  

\newcommand{\glb}{\left(}  
\newcommand{\grb}{\right)}  
\newcommand{\glc}{\left[}  
\newcommand{\grc}{\right]}  

\newcommand{\const}{\text{const}}
\newcommand{\TO}{,\ldots,}
\newcommand{\dd}[1]{\text{d}{#1\ }}   

\newcommand{\lD}{\lambda_{\text{T}}}

\newcommand{\mf}{\text{mf}}
\newcommand{\wfigure}[2]
   {\begin{figure}[htbp]
   \centerline{
 \epsfbox{#1.eps} 
  }
   \caption
   {#2}
   \label{f:#1}
   \end{figure}}
  
\begin{document}

\title{Universal correlations and coherence in quasi-two-dimensional
trapped Bose gases}

\author{Markus Holzmann$^1$, Maguelonne Chevallier$^2$, Werner
Krauth$^2$ }
\affiliation{$^1$LPTMC,  Universit\'e Pierre et Marie Curie, 4 Place
Jussieu, 75005 Paris, France; and LPMMC, CNRS-UJF,  BP 166, 38042
Grenoble, France} 
\email{markus@lptmc.jussieu.fr}
\affiliation{$^2$CNRS-Laboratoire de Physique Statistique, Ecole Normale
Sup\'{e}rieure, 24 rue Lhomond, 75231 Paris Cedex 05, France}
\email{werner.krauth@ens.fr}
\date{\today}

\begin{abstract}
We study the quasi-two-dimensional Bose gas in harmonic traps at
temperatures above the Kosterlitz--Thouless transition, where the
gas is in the normal phase. We show that mean-field theory takes into
account the dominant interaction effects for experimentally relevant
trap geometries. Comparing with Quantum Monte Carlo calculations,
we quantify the onset of the fluctuation regime, where correlations
beyond mean-field become important.  Although the density profile
depends on the microscopic parameters of the system, we show that
the correlation density (the difference between the exact and the
mean-field density) is accurately described by a universal expression,
obtained from classical-field calculations of the homogeneous strictly
two-dimensional gas. Deviations from universality, due to the finite
value of the interaction or to the trap geometry, are shown to be small for
current experiments.  We further study coherence and pair correlations
on a microscopic scale.  Finite-size effects in the off-diagonal density
matrix allows us to characterize the cross-over from Kosterlitz--Thouless
to Bose--Einstein behavior for small particle numbers. Bose--Einstein
condensation occurs below a characteristic number of particles which
rapidly diverges with vanishing interactions.
\end{abstract}

\maketitle

\section{Introduction}
In recent years, several experiments  \cite{Dalibard_2006,Clade}
studied two-dimensional ultra-cold atomic gases from the normal phase
down in temperature to the Kosterlitz--Thouless transition \cite{KT}
and into the low-temperature superfluid phase.  The interference  of two
simultaneously prepared two-dimensional gases evidenced the presence
of vortices \cite{Dalibard_2006}. Related experiments investigated
interaction and correlation effects \cite{Dalibard_2007,Clade} in the
density profile and in coherence patterns. For a quantitative description
of the Kosterlitz--Thouless transition and of the interaction effects,
it proved necessary to account for the quasi-two-dimensional nature of
the gas, that is to include thermal excitations in the strongly confined
$z$-axis in addition to the weak trapping potential in the $xy$-plane
\cite{Holzmann_Krauth_07,HCK}.

In weakly interacting two-dimensional Bose gases, the Kosterlitz--Thouless
phase transition occurs at relatively high phase-space density (number
of atoms per phase-space cell $ \lD^2=2 \pi \hbar^2/mT$).  This density
is $n_c \lD^2 \simeq \log(\xi_n/\gtilde)$ \cite{Hohenberg,TKTB,
Svistunov2D}, where $\gtilde$ characterizes the two-dimensional
interaction strength, $T$ is the temperature, $m$ the mass of the atoms,
and $n$ the density. The coefficient $\xi_n =380 \pm 3$ was determined
numerically using classical-field simulations \cite{Svistunov2D}. For the
ENS experiment of Hadzibabic et al. \cite{Dalibard_2006}, the critical
phase-space density is $n_c\lD^2 \sim 8$ in the center of the trap,
whereas in the NIST experiment of Clad\'e et~al. \cite{Clade}, $n_c\lD^2$
is close to $10$.  For phase-space densities between one and the critical
number, the gas is quantum degenerate yet normal. The atoms within
one phase-space cell are indistinguishable. They lose their particle
properties and acquire the characteristics of a field. The mean-field
description of particles interacting with a local atomic density $n(r)$
may further be modified through correlations and fluctuations.  Quantum
correlations can be several times larger than the scale $\lD$. This
gives rise to \quot{quasi-condensate} behavior inside the normal phase.

In this paper, we study the quantum-degenerate regime at high
phase-space density in the normal phase.  We first discuss the peculiar
quasi-two-dimensional thermodynamic limit where, as the number of
particles in the gas is increased, the interactions and the lattice
geometry are scaled such that a finite fraction of all particles are
in the excited states of the system. In this thermodynamic limit, the
Kosterlitz--Thouless transition takes place at a temperature comparable
to the Bose--Einstein transition temperature in the non-interacting case,
and the local-density approximation becomes exact.  We first clarify
the relation between different recent versions of quasi-two-dimensional
mean-field theory \cite{Dali,BBB, HCK} in the local-density approximation
(LDA), and also determine the finite-size corrections to the LDA.
We compare mean-field theory to a numerically exact solution obtained by
path-integral Quantum Monte Carlo (QMC) calculations with up to $N \gtrsim
10^5$ interacting particles in a harmonic trap with parameters chosen
to fit the experiments. We concentrate on the correlation density, the
difference between the exact density and the mean-field density at equal
chemical potential, and show that it is essentially a universal function,
independent of microscopic details. Within classical-field theory, the
correlation density is obtained from a reparametrization of known results
for the strictly two-dimensional homogeneous {system \cite{Svistunov2D_2}.
The classical-field results hold for small interaction parameters $\gtilde
\to 0$, but our full QMC solution accounts for corrections.  We compute
the correlation density by QMC and show that it is largely independent
of the trap geometry, the temperature, and the interaction strength.

We also study off-diagonal coherence properties, and the density--density
correlation function of the quasi-two-dimensional gas. It is well known
that even at high temperature, bosonic bunching effects enhance the
pair-correlation function on length scales below $\lD$ which for the
ideal Bose gas approaches the characteristic value $2 n^2$ at vanishing
separation.  In our case,  interference in the $z$-direction reduces the
in-plane density fluctuations even for an ideal gas and within mean-field
theory, and the reduction of the pair correlations from $2n^2$ no longer
proves the presence of beyond-mean-field effects.

We finally discuss finite-size effects in the quasi-two-dimensional
Bose gas.  For the density profile, they are not very large, but we point
out their great role for off-diagonal correlations. The latter are 
responsible for a cross-over between the physics of Bose--Einstein
condensation at small particle number and the Kosterlitz--Thouless
physics for larger systems; both regimes are of relevance for current
experiments. This cross-over takes place at a particle number $N \sim
\gtilde^{-2}$ which grows very rapidly as the interaction in the gas
diminishes.

\section{System parameters and mean-field description}
\subsection{Quasi-two-dimensional thermodynamics}
\slabel{quasi_two_dimensional}
We consider $N$ bosons in a three-dimensional pancake-shaped harmonic
potential with parameters $\omega_x=\omega_y=\omega$ and $\omega_z
\gg \omega$ at inverse temperature $\beta=1/T$.  The $z$ variable is
separate from $x$ and $y$, and we denote the three-dimensional vectors as
$\rvecD=(\rvec,z)$, and write two-dimensional vectors as $\rvec=(x,y)$,
and $r=|\rvec|$.

The quasi-two-dimensional regime of the Bose gas \cite{HCK} is
defined through a particular thermodynamic limit $N \to \infty$, where
the temperature is a fixed fraction $t\equiv T/\TBECZERO$ of the
Bose--Einstein transition temperature of the ideal two-dimensional
Bose gas, $\TBECZERO = \sqrt{6N}\hbar \omega/\pi$. Although the
two-dimensional trapped Bose gas undergoes a Bose--Einstein transition
only for zero interactions, $\TBECZERO$ still sets the scale for the
Kosterlitz--Thouless transition in the interacting gas\cite{TKTB,HCK}. In
the quasi-two-dimensional regime, a finite fraction of atoms remains in
excited states in $z$. The excitation energy is scaled as $\hbar \omega_z
\propto \TBECZERO$, which implies that $\omega_z$ increases as $N^{1/2}$
in the thermodynamic limit.

Interatomic collisions are intrinsically three-dimensional. Here,
we consider the experimentally relevant case where the range of the
scattering potential $r_0$ is much smaller than the typical lateral
extension $l_z=(m \omega_z/\hbar)^{-1/2}$, and also where $r_0$ is much
smaller than the inter-particle distances.  The interactions are then
described by the three-dimensional $s$-wave scattering length $a_s$,
and one may characterize the quasi-two-dimensional gas through a bare
effective two-dimensional interaction strength, $\gtilde$,
\begin{align}
\gtilde &= \frac{m g}{\hbar^2} \int \dd{z} \glc \psi_0(z) \grc^4
\elabel{two_d_three_d_interaction}
\\
g&= \frac{ 4 \pi \hbar^2 a_s}{m}, 
\end{align}
where $\psi_0(z)$ is the unperturbed ground state of the confining
potential and $g$ is the usual three-dimensional coupling constant.
For a harmonic confinement, $\gtilde= \sqrt{8 \pi}  a_s/l_z$, and
$a_s/l_z$ must be kept constant in the thermodynamic limit to obtain a
fixed two-dimensional interaction strength.

Quasi-two-dimensional scattering amplitudes depend logarithmically on
energy, $\epsilon$, in terms of a universal function of $a_s/l_z$ and of
$\epsilon/\hbar \omega_z \sim T/\omega_z$, which are both kept constant
in the quasi-two-dimensional thermodynamic limit. The logarithmic
energy dependence yields small corrections of order $(a_s/l_z)^2$
\cite{dima2000,dima2001,stoof2008} to the bare interaction $\gtilde$.
They can be neglected in the following.

The scaling behavior in the quasi-two-dimensional limit 
corresponds to the following reduced variables:
\begin{equation}
\begin{array}{ll}
\rtilde=r/l_T &  \ztilde=z/l_z\\
t= T/\TBECZERO   & \omegaztilde = \hbar \omega_z/\TBECZERO\\
\ntilde = n \lD^2 & \gtilde=  m g/(\sqrt{2\pi} l_z \hbar^2), 
\end{array}
\elabel{quasi_two_dimensional_scaling}
\end{equation}
where $l_T=(T/m\omega^2)^{1/2}$ is the thermal extension in the plane.
The quasi-two-dimensional limit consists in taking $N \to \infty$, with
$t, \omegaztilde$ and $\gtilde$ all constant. In this limit, $l_T/\lD=t
\sqrt{3 N/\pi^3} \gg 1$, so that macroscopic and microscopic length
scales separate, and the scaling of the three-dimensional density,
$n_{3d}$, is at constant
\begin{equation} \ntilde_{3d}=n_{3d} \lD^2 l_z.
\end{equation} 
In reduced variables, the normalization condition $N= \int \dd{\rvec}
n(\rvec)=(l_T/\lD)^2  \int \dd{\rvectilde} \ntilde(\rvectilde)$ is
expressed as
\begin{equation}
\int_0^\infty \dd{\rtilde} \rtilde \ntilde(\rtilde) = \frac{\pi^2}{6 t^2}.
\elabel{normalization_reduced}
\end{equation}
The local-density approximation becomes exact in the quasi-two-dimensional
limit.

In the ENS experiment, $^{87}$Rb atoms are trapped at temperatures $T
\approx 50-100 \, \text{nK}$. The in-plane trapping frequencies are
$\omega/(2\pi) \approx 50 \text{Hz}$ whereas the confinement is of order
$\omega_z/(2\pi) \approx 3 \text{kHz}$.  With $N \sim 2  \cdot 10^4$
atoms trapped inside one plane, typical parameters are $\TBECZERO \approx
300 \text{nK}$ (using $\hbar/k_B \simeq 7.64  \cdot 10^{-3} \text{nKs}$),
so that $\omegaztilde \approx 0.44-0.55$. The scattering length $a_s=5.2
\text{nm}$ leads to an effective coupling constant $\gtilde=0.13$,
using $\hbar/m\simeq 6.3 \cdot 10^{-8} m^2 s^{-1} A^{-1}$, where $A$
is the atomic mass number. In the NIST experiment, sodium atoms at
$T \approx \TBECZERO \approx 100 \text{nK}$ are confined by harmonic
trapping potentials with $\omega/(2\pi) \approx 20 \text{Hz}$, and
$\omega_z/(2\pi) \approx 1 \text{kHz}$. This is described by reduced
parameters $\gtilde=0.02$ and $\omegaztilde =0.50$. The critical
densities are $\ntilde_c \approx \log(380/\gtilde) \approx 8.2$ for
the ENS parameters, somewhat lower than the NIST value $\ntilde_c
\approx 9.9$.  Using the quasi-two-dimensional mean-field estimates
of Ref.~\cite{HCK}, the Kosterlitz--Thouless temperatures are located
at $t_{\KT}\equiv T_{\KT}/\TBECZERO \approx 0.69$ and $t_{\KT} \approx
0.74$, respectively.

The quasi-two-dimensional limit describes a kinematically two-dimensional
gas, whose extension in the $z$ direction is of the order of the
thermal wavelength $\lD$. As $\omegaztilde \equiv \lD^2/(2\pi t
l_z^2)$ is decreased, a system at finite $N$ turns three-dimensional.
This is already the case for the ideal quasi-two-dimensional gas (with
$\gtilde=0$) where the Bose--Einstein transition temperature crosses
over from two-dimensional to three-dimensional behavior as a function
of $\omegaztilde$, with asymptotic behavior given by
\begin{align}
t_\BEC \sim
\begin{cases}
 \glc \frac{\zeta(2)}{\zeta(3)}
\grc^{1/3}\omegaztilde^{1/3}-\frac{1}{6}\frac{\zeta(2)}{\zeta(3)}
\omegaztilde &\text{for $\omegaztilde \ll 1$} \\
1 - \frac{1}{ 2 \zeta(2)^{3/2}} \expb{-\omegaztilde}   
           &\text{for $\omegaztilde \gg 1$}
\end{cases}
\elabel{expansion_of_t}
\end{align}
(see \cite{HCK}).  In \eq{expansion_of_t}, the first term for
$\omegaztilde \ll 1$ describes three-dimensional Bose--Einstein
condensation in an anisotropic trapping potential.

For the interacting Bose gas, the nature of the Kosterlitz--Thouless
transition in two dimensions differs from the Bose--Einstein transition of
the three-dimensional gas.  For large $\omegaztilde$, universal features
of the Kosterlitz--Thouless transition are preserved, but the density
profiles and the value of the Kosterlitz--Thouless  transition temperature
depend on $\omegaztilde$ and $\gtilde$ \cite{HCK,Holzmann_Krauth_07}.
For small confinement strength $\omegaztilde$, a dimensional cross-over
between the two-dimensional Kosterlitz--Thouless transition and the
three-dimensional Bose--Einstein condensation takes place at particle
numbers such that the level spacing in the confined direction is
comparable to the (two-dimensional) correlation energies, $ t \omegaztilde
\lesssim \gtilde \ntilde/\pi$.

The quasi-two-dimensional limit differs from the \quot{experimentalist's}
thermodynamic limit where the atom number is increased in a fixed
trap geometry, and at constant temperature. In this situation,
the ratio between the microscopic and the macroscopic length scales,
$l_T/\lambda_T=T/(\hbar \omega\sqrt{2\pi})$, remains constant and finite.
The number of particles in any region of nearly constant density
remains also finite so that, in contrast to the quasi-two-dimensional
thermodynamic limit, corrections to the LDA persist.

\subsection{$N$-body and mean-field Hamiltonians}
\slabel{mean_field_hamiltonian}

The gas specified in \sect{quasi_two_dimensional} is described by the
Hamiltonian
\begin{align}
H=&H_0+V, \\ H_0=&
\sum_{i=1}^N \glc- \frac{\hbar^2\nabla^2_i}{2m} 
+ \frac{1}{2} m \glc \omega^2 \rvec_i^2
+ \omega_z^2 z_i^2 \grc\grc,  \elabel{non_interacting_hamiltonian}
\\
V=& \sum_{i<j=1}^N v(|\rvecD_i-\rvecD_j|), \elabel{n_body_interaction}
\end{align}
where $v$ is the three-dimensional interaction potential. We compute the
$N$-body density matrix at finite temperature using three-dimensional
path-integral QMC methods.  We thus obtain all the thermodynamic
observables \cite{qmc,mchf,Holzmann_Krauth_07} for up to $N=10^6$. QMC
calculations have clearly demonstrated the presence of a Kosterlitz--Thouless
transition \cite{HCK} for parameters corresponding to the ENS experiment.

In the mean-field approximation, one replaces the $N$-body interaction
between atoms in \eq{n_body_interaction} by an effective single-particle
potential. The mean-field Hamiltonian writes
\begin{equation}
H_\mf = H_0 + V_\mf,
\elabel{mean_field_hamiltonian}
\end{equation}
where 
\begin{equation}
V_\mf =  \sum_{i=1}^N 2 g n_{3d}(\rvecD_i)-g \int \dd{\rvecD} [n_{3d}(\rvecD)]^2
\elabel{mean_field_interaction}
\end{equation}
is the mean-field potential energy. From the corresponding
partition function in the canonical or grand-canonical ensemble, all
thermodynamic quantities can be calculated. The three-dimensional density
$n_{3d}(\rvecD)$ inside the mean-field potential must be determined
self-consistently. In all situations treated in the present paper,
self-consistency is reached through straightforward iteration.

Mean-field theory leads to an effective Schr\"odinger equation for
the single-particle wavefunction, $\Psi_j(\rvecD)$, of energy $\epsilon_j$,
\begin{multline}
\left[ -\frac{\hbar^2 \nabla_{\rvec}^2}{2m} + \frac12 m \omega^2 r^2
-\frac{\hbar^2 \partial^2}{2m\partial z^2} + \frac12 m \omega_z^2 z^2
+ 2 g n_{3d}(\rvecD)
 \right] 
\\
 \Psi_j(\rvecD)
=
 \epsilon_j  \Psi_j(\rvecD)
 \elabel{mean-field3D}
\end{multline}
together with the total density
\begin{equation}
n_{3d}(\rvecD) = \sum_i \frac{\Psi_j^*(\rvecD) \Psi_j(\rvecD)}{e^{\beta(\mu-\epsilon_j)}-1}.
\elabel{mf3Ddensity}
\end{equation}
The exact solution of the mean-field eigenfunctions and eigenvalues for
finite systems, is rather involved, but considerably simplifies in the
local-density approximation.

At finite $N$, in the canonical ensemble, we solve the mean-field
equations through a Quantum Monte Carlo simulation with $N$ particles
which avoids an explicit calculation of all eigenfunctions.  In contrast
to the usual interaction energy within QMC, consisting, in general,
of a pair interaction potential, the mean-field interaction energy
is simply given in terms of an anisotropic, single-particle potential
proportional to the three-dimensional density profile, $n_{3d}(\rvecD)$,
as in \eqtwo{mean_field_hamiltonian}{mean_field_interaction}. This
interaction potential must be obtained self-consistently as usual in
mean-field.  Once self-consistency in the density is reached, one can
compute correlation functions and off-diagonal elements of the reduced
one-body density matrix.

\subsection{Mean-field: Local-density approximation}

Mean-field theory simplifies in the quasi-two-dimensional thermodynamic
limit, as the local-density approximation then becomes exact. This is
because the natural length scale of the system, $\lD$,  separates from the
macroscopic scale $l_T$  of variation of the density ($\lD/l_T \to 0$).
The particle numbers inside a region of constant density diverges. The
decoupling of length scales implies that the $xy$-dependence of the single
particle wavefunctions in \eq{mean-field3D} separates in the thermodynamic
limit. Using scaled variables, \eq{quasi_two_dimensional_scaling},
this yields
\begin{multline}
\glc -\frac{1}{2}\frac{d^2}{d\ztilde^2}
+ \frac{1}{2}\ztilde^2 
    + \frac{2 \gtilde t }{\sqrt{2 \pi} \omegaztilde} 
           \ntilde_{3d}^{\mf}(\rtilde,\ztilde)
      \grc  \\  \phitilde_{\nu}(\rtilde,\ztilde)=
        \frac{\epstilde_{\nu}(\rtilde)}{ \omegaztilde} \phitilde_{\nu}(\rtilde,\ztilde),
\elabel{schroedinger_reno}
\end{multline}
for the eigenfunctions, $\phitilde_{\nu}(\rtilde,\ztilde)$, and eigenvalues, 
$\epstilde_{\nu}(\rtilde)$, in the confined direction,
at a given radial distance, $\rtilde$.
The reduced local density $\ntilde_{3d}^{\mf}$ is given by
the normalized wavefunctions $\phitilde(\rtilde,\ztilde)$:
\begin{align}
\ntilde_{3d}^{\mf}(\rtilde,\ztilde) &= \sum_\nu \phitilde_\nu^2(\rtilde,\ztilde) 
\ntilde_\nu^{\mf}(\rtilde)
\nonumber
\\
\ntilde_\nu^{\mf} (\rtilde) &= - \logc{1 - \expb{\mutilde(\rtilde) - \epstilde_\nu(\rtilde)/t}}.
\elabel{ntilde_nu_finite}
\end{align}
The position-dependence in \eq{schroedinger_reno} and
\eq{ntilde_nu_finite} only enters parametrically through the
$\rtilde$-dependence of the chemical potential,
\begin{equation}
\mutilde(\rtilde) = \mutilde - \frac{\rtilde^2}{2}, 
\elabel{shift_in_mu}
\end{equation}
and the local-density approximation becomes exact in the
quasi-two-dimensional thermodynamic limit.  Within LDA, density profiles
(as in \fig{density_r_inset_071}) are directly related to the equation
of state $\ntilde(\mutilde)$ of a quasi-two-dimensional system, which
is homogeneous in the $xy$-plane.

The Schr\"{o}dinger equation of \eq{schroedinger_reno} is conveniently
written in the basis $\SET{\psi_0,\psi_1,\dots,\psi_n,\dots}$ of the
one-dimensional harmonic oscillator with $\omega=m=1$, as it diagonalizes
\eq{schroedinger_reno} for $\gtilde=0$. Using
\begin{equation}
\phitilde_\nu(\ztilde)= \sum_\mu a_{\mu \nu} \psi_\mu(\ztilde),  
\elabel{phi_psi}
\end{equation}
(where we have dropped the index corresponding to $\rtilde $ or,
equivalently, to  $\mutilde$), we can write it as a matrix equation
\begin{equation}
\glb \Avec- \frac{\epstilde_\nu}{\omegaztilde} \grb\avec_\nu=0
\elabel{amat1}
\end{equation}
with eigenvalues $\epstilde_\nu/\omegaztilde$ and eigenvectors
$\avec_\nu=\SET{a_{0 \nu} \TO a_{n \nu}}$, of the $(n+1)\times (n+1)$
matrix
\begin{equation}
\Avec_{\mu\nu} =  \nu \delta_{\mu \nu} + 
\frac{2 \gtilde t}{\sqrt{2 \pi} \omegaztilde} 
\int \dd{\ztilde}  \psi_\mu(\ztilde)
\ntilde_{3d}^{\mf}(\mutilde,\ztilde) \psi_\nu(\ztilde)
\elabel{amat}
\end{equation}
(where $0 \le \mu,\nu \le n$) and the density
$\ntilde_{3d}^{\mf}(\mutilde,\ztilde)=-\sum_\nu \phitilde_\nu^2(\ztilde)
\log[1-\exp(\mutilde-\epstilde_\nu/t)]$.  The wavefunctions $\psi_\nu$
are easily programmed (see, e.g., \cite{SMAC} sect. 3.1), and the
self-consistent mean-field solutions at each value of $\mutilde$ can be
found via iterated matrix diagonalization.

This full solution of the LDA mean-field equations is analogous to the
one in Ref.\cite{Dali}. The mean-field version used in \cite{BBB},
however, neglects  the off-diagonal couplings in $A_{\nu\mu}$ with
$\nu \ne \mu$.  In \cite{HCK}, we used a simplified mean-field
potential in order to reach explicit analytical expressions.
These different mean-field approximations essentially coincide at all
relevant temperatures \cite{BBB}, but ground-state occupations in $z$
slightly differ.  Further replacing the coupling constant $\tilde{g}$
by $\tilde{g} \tanh^{1/2}(\tilde{\omega}_z/2t)$ has allowed us, in
Ref.~\cite{HCK}, to improve the  agreement with the QMC results close
to the transition. This is because mean-field theory overestimates the
effect of the interactions in the fluctuation regime.  In the following
we always quantify beyond-mean-field corrections with respect to the
full LDA solution of \eq{amat}.

\section{Correlation density and universality}
\slabel{correlation_effects}

Comparisons between the QMC and the mean-field density profiles
are shown in \fig{density_r_inset_071} for the ENS parameters at
reduced temperature $t=0.71$, slightly above the Kosterlitz--Thouless
temperature.  Finite-size effects as well as deviations from mean-field
theory are visible for $n \lD^2 \gtrsim 5$.  In this section,
we concentrate on correlation corrections to mean-field theory in the
thermodynamic limit and postpone the discussion of finite-size effects to
\sect{finite-size-effects}. We analyze the QMC density
profiles within the validity of the LDA, \eq{shift_in_mu}, and compare QMC
and mean-field densities at the same local chemical potential \footnote{In
general, at equal chemical potential, the  mean-field density differs
from the exact one, and so does also the total number of particles in
the trap.} which defines the correlation density, $\Delta \ntilde$,
\begin{equation}
\Delta \ntilde(\mutilde)= \ntilde(\mutilde)-\ntilde_{\mf}(\mutilde).
\end{equation}  
As in experiments, the chemical potential is not a control parameter of
the QMC calculation, but it can be obtained from a fit of the wings of
the density profile with $\ntilde \lesssim 1$ to the mean-field equation
of state.

Mean-field effects take into account the dominant interaction effects
which, in particular, determine shape and energies of the ground and
excited states in the tightly confined direction. One expects that
correlation effects do not modify these high-energy modes, but merely
affects the low-energy distribution of $xy$-modes inside the confining
ground state, $\phitilde_0(\ztilde)$.  This assumption is supported
by a direct comparison of the normalized density distribution in
$\ztilde$ between the QMC solution and the mean-field approximation
(inset of \fig{density_r_inset_071}) at different radial distances,
$\rtilde$.  For small $\rtilde$, the ground state of the confining
potential is strongly populated. For larger $\rtilde$, higher modes
of the one-dimensional harmonic oscillator are thermally occupied,
and the density distribution broadens. However, the normalized density
profile in $\ztilde$ is everywhere well described by mean-field theory,
and correlation effects hardly modify the mode structure in the confined
direction.

\wfigure{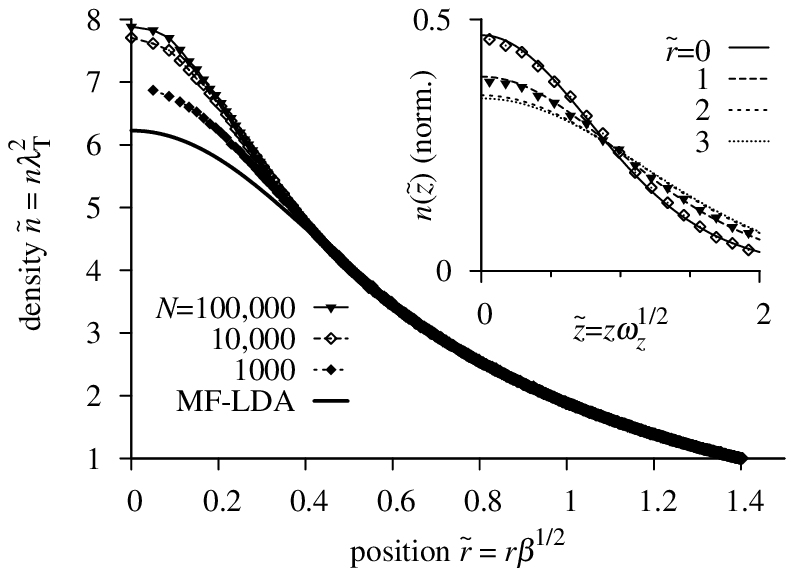}{QMC density profile for the ENS parameters
($\gtilde=0.13$, $\omegaztilde=0.55$) at temperature $t=T/\TBECZERO =
0.71$ for different values of the particle number, compared to the
LDA mean-field solution of \sect{mean_field_hamiltonian} at 
the same total number of particles. The inset shows
the density distribution $n(\ztilde)$ at different values of the radial
distance $\rtilde$.}

In the homogeneous two-dimensional gas, corrections to mean-field
theory at small $\gtilde$ are described by classical-field theory, and
correlation effects in the density profile can be expressed in terms of a
universal function of $\beta(\mu-2 g n_{\mf})$ \cite{Svistunov2D_2,TKTB}.
In a quasi-two-dimensional geometry,  the corresponding relevant quantity
is given by the local mean-field gap $\Delta_{\mf}$ between the local
ground-state energy in the confining potential and the local chemical
potential,
\begin{equation}
\Delta_{\mf}(\rtilde)=\epstilde_0(\rtilde)/t-\mutilde(\rtilde).
\end{equation}
Within mean-field theory, it fixes the local $xy$-density in the
ground state of the confining potential 
\begin{equation}
\ntilde_0^{\mf}(\rtilde)=-\log \glb1-\expa{-\Delta_{\mf}(\rtilde)} \grb.
\elabel{dens_gap}
\end{equation}
In the strictly two-dimensional limit, we have 
\begin{equation}
\Delta_{\mf} \to \beta( 2
\gtilde \hbar^2 n_{\mf}/m-\mu)= \gtilde \ntilde_{\mf}/\pi-\mutilde, 
\elabel{2d_gap}
\end{equation}
where
$n_{\mf}$ is the total mean-field density. Deviations  are noticeable
for large densities, as  illustrated in \fig{LDA_analysis_0.71_013}
(we have absorbed the zero-point energy $\omegaztilde/(2t)$  in the
chemical potential).  Within LDA, we expect that the correlation
density $\Delta \ntilde$ coincides to leading order in $\gtilde$
with the classical-field-theory results of the homogeneous strictly
two-dimensional system \cite{Svistunov2D_2}, expressed as functions of
the mean-field gap $\Delta_{\mf}$ \footnote{Similar to the mean-field
gap, one may introduce an effective mean-field coupling constant
$\gtilde_{\mf}=\gtilde \int \dd{ \ztilde} |\phitilde_0(\ztilde)|^4/
\int \dd{\ztilde} |\phitilde_0(\ztilde)|^2$ which accounts for
modifications of the in-plane interactions in the ground state of the
confining potential. This leads to small corrections, not visible for
the experimental parameters considered in this paper.}.

\wfigure{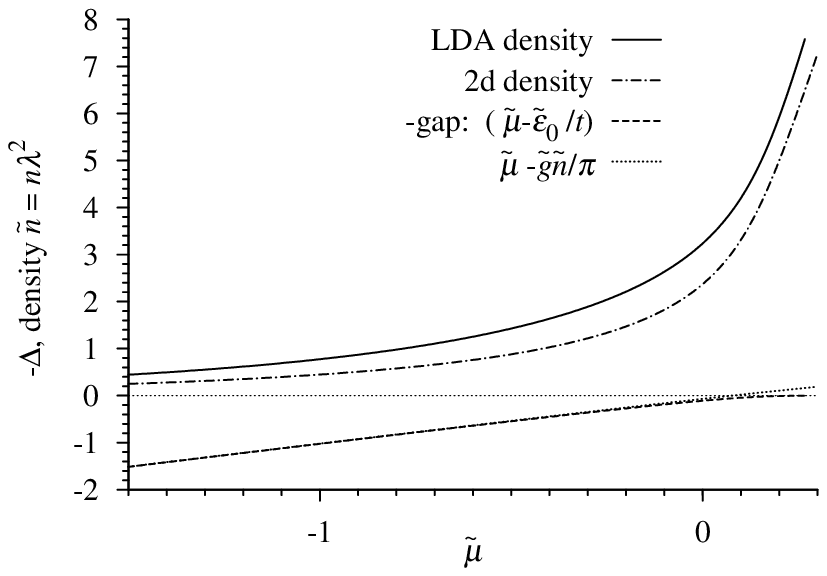}{Mean-field equation of state and gap
$\Delta_{\mf}$ for ENS parameters $\gtilde=0.13$, $\omegaztilde=0.55$,
$t=0.71$. The gap differs from the approximation $\mutilde-
\gtilde\ntilde/\pi$ only at high density. The 2d mean-field curve (see
\eqtwo{dens_gap}{2d_gap}) illustrates the dependence of the equation of
state on microscopic parameters.}

In Ref.~\cite{Svistunov2D_2}, the critical density $\ntilde_c$ and
chemical potential $\mutilde_c$ at the Kosterlitz--Thouless transition
were determined to
\begin{align}
\ntilde_c &= \log \frac{\xi_n}{\gtilde}, \quad \xi_n =380 \pm 3,\elabel{ntilde_mf} \\
\mutilde_c &=
\frac{\gtilde}{\pi} \log \frac{\xi_\mu}{\gtilde}, \quad \xi_\mu=13.2 \pm 0.4,
\end{align}
and the equation of state in the neighborhood of the transition was written as
\begin{equation}
\ntilde-\ntilde_c=2\pi\lambda (X), \text{with $X= (\mutilde-\mutilde_c)/\gtilde$}.
\elabel{diff_density}
\end{equation}
The function $\lambda(X)$ was tabulated.
Consistent with the classical-field approximation,
we can expand \eq{dens_gap} to leading order in $\Delta_{\mf}$, 
\begin{equation}
\ntilde^{\mf}_0(\Delta_{\mf}) = -\log(\Delta_{\mf}) \elabel{ncl_mf}, 
\end{equation}
so that we can express $X$ through $\Delta_\mf$:
\begin{multline}
X(\Delta_{\mf}) = 
\frac{\ntilde_\mf }{\pi} - \frac{\Delta_\mf}{\gtilde} - \frac{\mutilde_c}{\gtilde} \\ 
= - \glc \frac{\Delta_\mf}{\gtilde} + 
\frac{1}{\pi} \logb{\xi_\mu \frac{\Delta_{\mf}}{\gtilde}} \grc.
\elabel{re_param}
\end{multline}
Thus, we obtain the correlation density as a function of $\Delta_\mf$:
\begin{equation}
\Delta\ntilde=2 \pi \lambda[X(\Delta_{\mf})] + \logb{\xi_n \Delta_\mf/\gtilde}, 
\elabel{correlation_density}
\end{equation}
and a straightforward inversion of \eq{re_param} allows us to translate
the data of \cite{Svistunov2D_2} in order to obtain the correlation
density as a function of the mean-field gap.  An empirical interpolation
of the numerical data with $\lesssim 10 \%$ error inside  the
fluctuation regime, $\Delta_{\mf}^c \le \Delta_{\mf}\le \Delta_{\mf}^f$
($\Delta_{\mf}^c$ and $\Delta_{\mf}^f$ are defined below), is given by
\begin{equation}
\Delta \ntilde(\Delta_{\mf}/\gtilde) \simeq \frac{1}{5}\glb -1 +
\frac{1}{\Delta_{\mf}/\gtilde}  \grb \frac{1}{1+\pi
\Delta_{\mf}^2/\gtilde^2}
\elabel{deltan_fit}
\end{equation}
where positivity is imposed since, within classical field
theory, the correlation density must be positive and of order
$(\Delta_{\mf}/\gtilde)^{-2}$ for $\Delta_{\mf}/\gtilde \to \infty$.

\wfigure{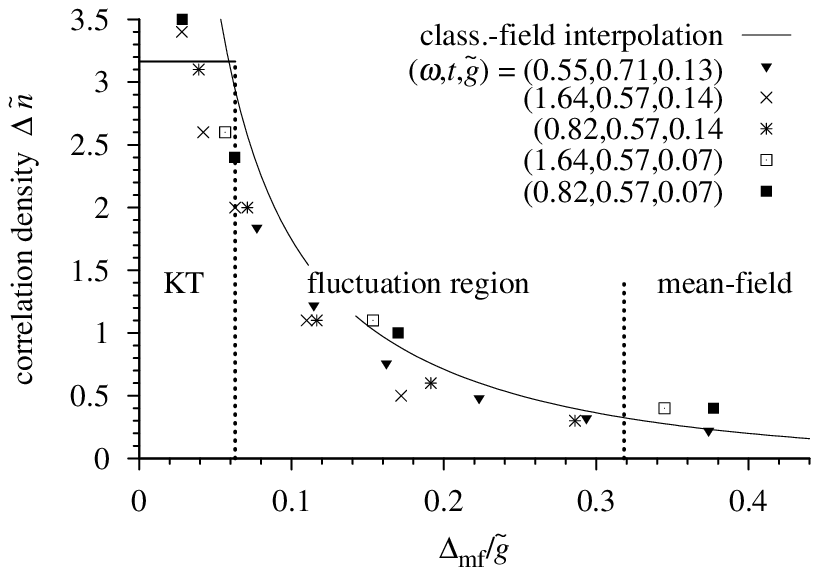}{Correlation density $\Delta \ntilde$
vs. rescaled mean-field gap $\Delta_{\mf}/\gtilde$. QMC data for various
interaction strengths and confinements $\omegaztilde$ are compared
with the interpolation \eq{deltan_fit} of classical-field results
\cite{Svistunov2D_2}.}

In \fig{test_universal}, we plot the classical-field results for the
correlation density as a function of the rescaled mean-field gap.
We also indicate the onset of the Kosterlitz--Thouless transition
($\mutilde=\mutilde_c$, $X=0$ in \eq{correlation_density}) at
$\Delta_\mf^c/\gtilde = 0.0623 $, which yields $\Delta\ntilde_c = 3.164$.
The correlation density in the normal phase is thus finite for all
interactions, whereas the mean-field density  diverges  as $\ntilde_{\mf}
=-\log \Delta_{\mf}^c \propto -\log g$ for small interactions at $T_\KT$.
In \fig{test_universal}, we furthermore compare the classical-field
data for the correlation density with the results of QMC simulations
of quasi-two-dimensional trapped Bose gases with different coupling
constants $\gtilde$ and confinement strengths $\omegaztilde$.
 The QMC data illustrates that the external trapping
and the quasi-two-dimensional geometry preserve universality in the
experimental parameter regime.  However, the finite coupling constant
$\gtilde$ introduces small deviations due to quantum corrections.

From \fig{test_universal}, we further see that the correlation density
is reduced to roughly $10\%$ of its critical value for mean-field
gaps $\Delta_{\mf}^f \simeq  \gtilde/\pi$. Thus, only densities
with $\ntilde \gtrsim \ntilde_f \approx \ntilde_{\mf}(\Delta_{\mf}
\approx \gtilde/\pi)$ are significantly affected by correlations, and
$\ntilde_f$ can be considered as the boundary of the fluctuation regime.
In fact, perturbation theory fails inside this regime.  For a
strictly two-dimensional system, we have 
\begin{equation} \ntilde_f
\approx \log(\pi/\gtilde),  
\end{equation} 
and the fluctuation regime is reached for densities $\ntilde \gtrsim
\ntilde_f$. Outside  the fluctuation regime, $\ntilde \lesssim \ntilde_f$,
mean-field theory is rather accurate, and can be improved perturbatively,
if necessary.

To understand this criterion, which is important for the
Kosterlitz--Thouless to Bose--Einstein cross-over at small $N$ (see
\sect{finite-size-effects}), we briefly analyze the perturbative structure
of the two-dimensional single-particle Green's function beyond mean-field
theory \cite{TKTB}. Within classical-field theory, second-order diagrams
are ultraviolet convergent.  Each additional higher order brings in a
factor $\hbar^2 \gtilde/m$ for the interaction vertex, one integration
over two-momenta, a factor $T$, and two Green's functions (the internal
lines). Dimensional analysis of the integrals involved shows that each
vertex insertion adds a factor $\gtilde/\Delta_{\mf}$. This implies that
perturbation theory fails for $\gtilde/\Delta_{\mf} \gtrsim 1$.

For lower densities, $1 \lesssim \ntilde \lesssim \ntilde_f$, the gas
is quantum degenerate, yet it is accurately described by mean-field
theory. In contrast to fully three-dimensional gases, the
quantum-degenerate regime can be rather broad in two dimensions
for gases with $\gtilde \ll 1$.  In this regime, the density, yet
normal, is no longer given by a thermal Gaussian distribution. This
was observed in the NIST experiment \cite{Clade} where $\ntilde_f
\simeq 5$.  In \fig{density_NIST_075}, we illustrate this effect via
the approximations for the tails of the distribution $\ntilde(\rtilde)$
with between one and five Gaussians, as
\begin{equation}
\ntilde(\rtilde)
=\frac{\pi^2}{6t^2} \sum_{k=1}^{k_{\max}} k \pi_k \expc{- \frac{ k \rtilde^2}{2}}.  
\elabel{gaussian_fit} 
\end{equation}
where $\pi_k$ is determined by the formal expansion of the logarithm
in \eq{ntilde_nu_finite}, but also appears as a cycle weight  in the
path-integral representation of the bosonic density matrix where they
can be measured (see the inset of \fig{density_NIST_075}). The successive
approximations have no free parameters.

\wfigure{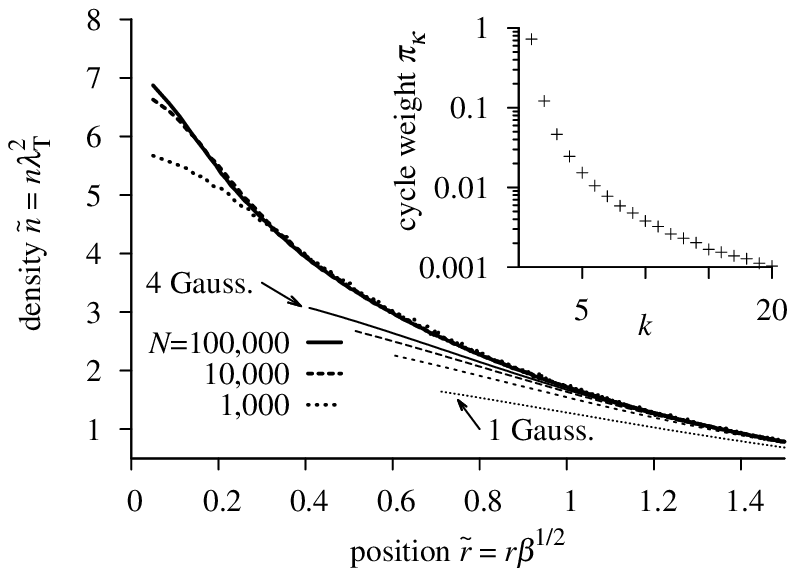}{QMC density profile for the NIST parameters
$t=T/\TBECZERO = 0.75$, $\gtilde=0.02$, and $\omegaztilde=0.5$,
for different $N$. The data are compared to the expansion of
\eq{gaussian_fit}, considering the largest ($k=1$) and the four largest
terms ($ k =1 \TO 4$). The inset shows the QMC cycle weights $\pi_k$
for $N=100\,000$ (see \cite{HCK}).}

\figg{nc_r} summarizes the density profiles of a strictly two-dimensional
Bose gas in the limit $\gtilde \to 0$ where the classical-field
calculations determine the correlation density.  At the
critical temperature $T_\KT$, the density in the center of the trap
is critical, $\ntilde(0)=\ntilde_c$. Correlation effects are important
only in the fluctuation regime, $\rtilde \lesssim 1.25 \sqrt{\gtilde}$,
where $\ntilde(\rtilde) \gtrsim \ntilde_f$.  However, the distribution
of the correlation density introduces no further qualitative features to
the mean-field component. The density profile may be integrated using the
interpolation formula for the correlation density, \eq{deltan_fit}. The
critical temperature of the strictly two-dimensional Bose gas as a
function of the total number of particles is given by
\begin{equation*}
\frac{T_\KT}{\TBECZERO}
\simeq \left( 1+ \frac{3 \gtilde}{\pi^3} \log^2 \frac{\gtilde}{16} + 
\frac{ 6 \gtilde}{16 \pi^2} \left(15+ \log  \frac{\gtilde}{16}\right) \right]^{-1/2}.
\elabel{Tkt}
\end{equation*}
This expression includes correction of order $\gtilde \log \gtilde$
compared to the mean-field estimate of Refs~\cite{TKTB,HCK}.
Since corrections beyond classical-field theory are rather small (see
\fig{test_universal}), the Kosterlitz--Thouless temperature of the
strictly two-dimensional trapped Bose gas is accurately described by
this equation even for large coupling constants.

\wfigure{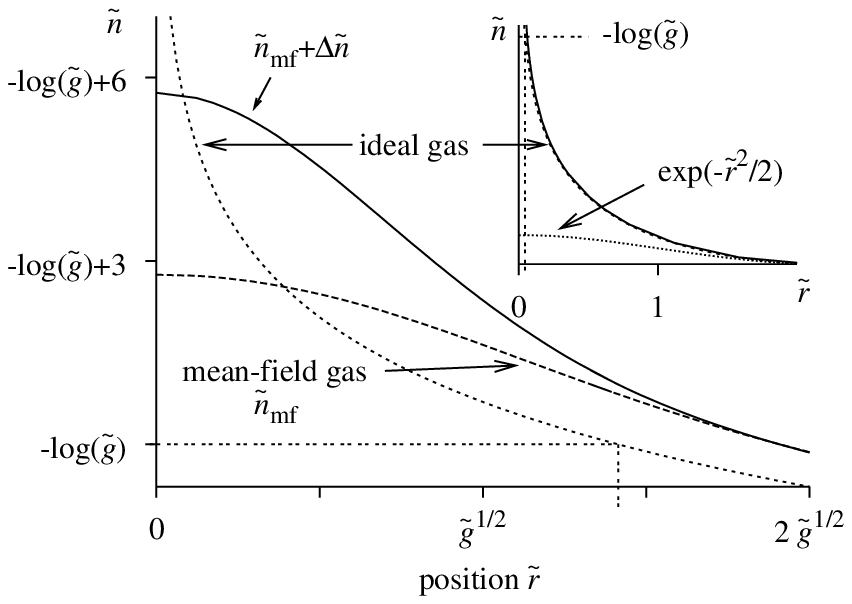}{Schematic density profile of a strictly two-dimensional
trapped Bose gas at $T_\KT$ for $\gtilde \to 0$.  For $\rtilde \gg
\sqrt{\gtilde}$, $\ntilde$ coincides with the ideal gas at $\TBECZERO$
(inset, the classical Boltzmann distribution $e^{-\rtilde2/2}$ is
given for comparison). In the fluctuation regime, for $\rtilde  \lesssim
1.25\sqrt{\gtilde}$, mean-field and correlation effects become important.
The density diverges as $ \sim \log(1/\gtilde)$, yet the correlation
contribution $ \Delta \ntilde$ remains finite.}

For general quasi-two-dimensional gases, \fig{nc_r} remains
qualitatively correct, but the LDA mean-field density profile in  the
quasi-two-dimensional geometry must be used.  Numerical  integration
of this $\gtilde \to 0$ density profile for the ENS parameters with
$\omegaztilde=0.55$  and $\gtilde=0.13$ leads $T_\KT \simeq 0.71 \,
\TBECZERO$, in close agreement with $T_\KT \simeq 0.70 \, \TBECZERO$
determined in Ref.~\cite{Holzmann_Krauth_07} directly from QMC
calculations using finite-size extrapolations. The quasi-two-dimensional
transition temperature is smaller than the one of the
strictly two-dimensional gas ($T^{2\text{d}}_\KT \simeq 0.86 \,
\TBECZERO$ for $\gtilde =0.13$).  For the NIST parameters ($\gtilde=0.02$,
$\omegaztilde=0.5$), we have $T_\KT \simeq 0.74 \, \TBECZERO$ from the
integration of the $\gtilde \to 0$ density, and QMC data indicate a
transition slightly below this value.

\section{Finite-size effects and Bose--Einstein cross-over}
\slabel{finite-size-effects}

\subsection{Central coherence}
\slabel{1body}
 
In the normal phase, the off-diagonal elements of the single-particle
density matrix remain short-ranged,  so that they can be described
locally.  From the self-consistent eigenfunctions of the mean-field
Schr\"odinger equation, \eq{mean-field3D} and \eq{mf3Ddensity}, we also
obtain the off-diagonal reduced single-body density matrix:
\begin{equation}
\ntilde^{(1)}_{\mf}(\rvecD;\rvecD')
= \lambda^2 l_z
\sum_j \frac{\Psi_j^*(\rvecD) 
\Psitilde_j(\rvecD')}{\expb{\mutilde-\beta\epsilon_j}-1}.
\end{equation} 
In the local-density approximation, we can separate the contributions of 
the different transverse modes, and we obtain
\begin{equation}
\ntilde^{(1)}_{\mf}(\rvecD;\rvecD')
=
\sum_{\nu}
\ntilde^{(1)}_{\mf,\nu}(\rvec;\rvec') \phitilde_\nu(\ztilde) \phitilde_\nu(\ztilde')
\elabel{nconf}
\end{equation}
with
\begin{equation}
\ntilde^{(1)}_{\mf,\nu}(\rvec;\rvec')
=
\int \frac{\text{d}^2\kvec}{(2\pi)^2}
\frac{\lD^2 \expa{i  \kvec \cdot (\rvec-\rvec')}}{\expa{\beta\hbar^2 k^2/2m 
+\Delta_{\mf}(\rtilde)}-1}.
\elabel{noff}
\end{equation} 
Here we have used that within the LDA, the density remains constant on
the scale $\lD$, so that the mean-field gaps at $\rtilde$ and $\rtilde'$
are the same.

At low densities, where the mean-field gap is large, $\Delta_{\mf}
\gg 1$, we can expand the Bose function in \eq{noff} in powers
of $\expb{-\Delta_{\mf}}$, and off-diagonal matrix elements rapidly
vanish for distances larger than the thermal wavelength $\lD$. At higher
densities, in the quantum-degenerate regime, $\Delta_{\mf} \ll 1$, many
Gaussians contribute, and coherence is maintained over larger distances.
In the limit $\Delta_{\mf} \to 0$, we can expand the denominator in
\eq{noff}, $\expc{\beta \hbar^2k^2/2m+\Delta_{\mf}} -1\approx \beta
\hbar^2k^2/2m+\Delta_{\mf} $, and the off-diagonal density matrix decays
exponentially. In this regime, the local mean-field coherence length is
given by  $\xi_{\mf}=\lD/\sqrt{4\pi \Delta_{\mf}}$.

In \fig{coh_ENS} and \fig{coh_NIST} we compare  the
normalized off-diagonal coherence function in the center of the trap
\begin{equation}
c(r)=\frac{\int \dd{z} n^{(1)}_{3d}(r,z;0,0)}{\int \dd{z} n^{(1)}_{3d}(0,z;0,0)}
\end{equation} 
from QMC calculations with LDA for the ENS and NIST conditions.  We see
that for $\ntilde \lesssim \ntilde_f$, as in the case of the density
profile, mean-field theory accurately describes the single-particle
coherence. However, it is evident that at higher densities, $\ntilde
\gtrsim \ntilde_f$, where correlation effects for the diagonal elements
of the density matrix are important, mean-field theory also fails to
describe the off-diagonal matrix elements.

To characterize the decay of the off-diagonal density matrix in
the fluctuation regime, $\ntilde \gtrsim \ntilde_f$, we consider a
simple one-parameter model which neglects the momentum dependence
of the self-energies in the ground state of the confining
potential. The single parameter of the model, the effective
local gap $\Delta(\rtilde)$, is chosen such that it reproduces
the local density of the QMC data.  The density matrix of this
\quot{gap}-model, $\ntilde^{(1)}_{\Delta}(\rvecD;\rvecD') = \sum_{\nu}
\ntilde^{(1)}_{\Delta,\nu}(\rvec;\rvec') \phitilde_\nu(\ztilde)
\phitilde_\nu(\ztilde')$, is a straightforward  generalization of
mean-field theory, where  in \eq{noff}, we replace
\begin{equation}
\Delta_\nu = 
   \begin{cases}
      \Delta \ & \text{for $\nu=0$}\\
      \Delta_\mf \ & \text{otherwise} 
   \end{cases} . 
\elabel{gap_model}
\end{equation}
To fix the gap $\Delta$ of this model, we  require that the diagonal
elements of the density matrix reproduces the exact density
\begin{equation}
\ntilde^{(1)}_{\Delta,0}(\rvec;\rvec) = \ntilde^{\mf}_0(\rtilde)+
\Delta \ntilde(\rtilde).
\end{equation}
 
\wfigure{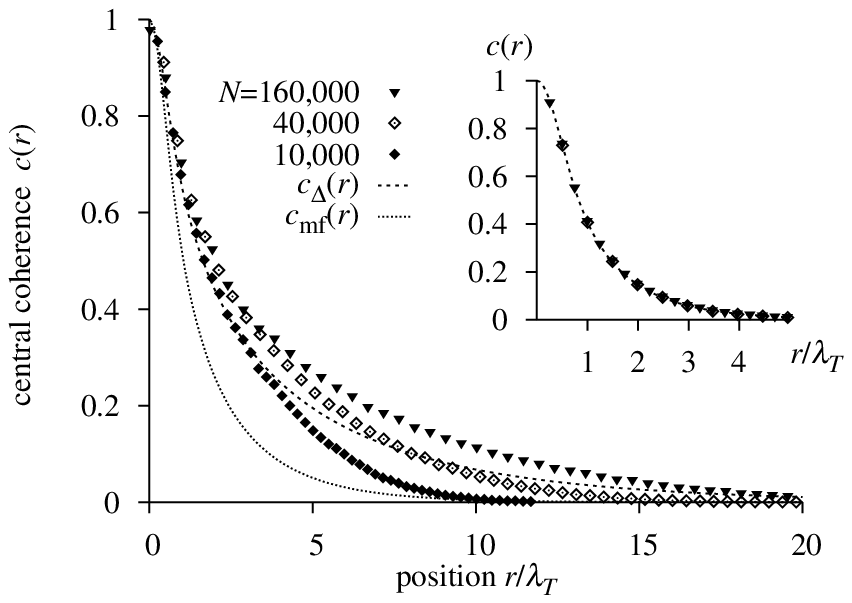}{ Off-diagonal coherence $c(r)$ for ENS parameters
with $t=0.71$ (main graph, $\ntilde > \ntilde_f$) and $t=0.769$
(inset, $\ntilde < \ntilde_f$) compared to the mean-field prediction
$c_{\mf}(r)$ and the gap model of \eq{gap_model}. In the fluctuation
regime, finite-size effects for off-diagonal correlations are more
pronounced than for the density (see \fig{density_r_inset_071}).}

Outside the fluctuation regime the gap model reduces to the mean-field
limit. Inside the fluctuation regime, where a direct comparison of the
coherence with mean-field theory is not very useful, the gap model
provides the basis to quantify off-diagonal correlations. It
cannot describe the build-up of quasi-long-range order at the
Kosterlitz--Thouless transition, but its correlation length $\xi_{\Delta}
=\lambda_T/\sqrt{4\pi \Delta(\rtilde)} > \xi_\mf$ bounds from below
the true correlation length in the normal phase. In \fig{coh_ENS},
we show that the gap model accounts for the increase of the coherence
length inside the fluctuation regime, $\ntilde > \ntilde_f$ for the
ENS parameters. For smaller interactions, as in the NIST experiment,
finite-size effects qualitatively  change the off-diagonal elements of
the density matrix (see \fig{coh_NIST}).

\wfigure{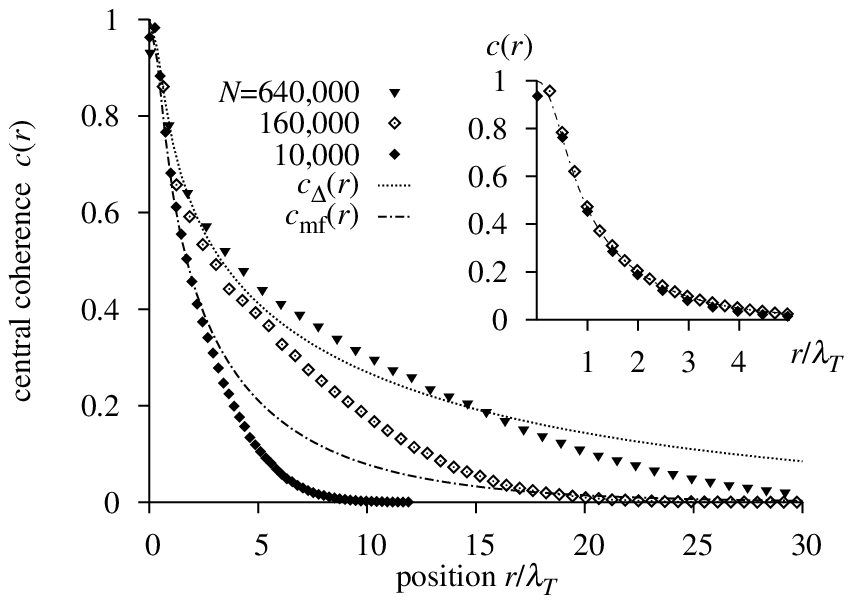}{Off-diagonal coherence
$c(r)$ for NIST parameters with $t=0.74$ (main graph) and
$t=0.769$ (inset) in comparison with the mean-field prediction,
$c_{\mf}(r)$, and the gap model, $c_{\Delta}(r)$, defined in
\eq{gap_model}.  At $t=0.769$, the total central density is  $\ntilde(0)
\simeq 5.1 < \ntilde_f$, and the system is  outside the fluctuation
regime.  At $t=0.74$, $\ntilde(0)\simeq 10.5 > \ntilde_f$, and the system
is close to the Kosterlitz--Thouless transition, $\Delta_{\mf}/\gtilde
\simeq 0.08$.  Strong finite-size effects are evident in the fluctuation regime.}

\subsection{Density profile}

Finite-size effects in the density profile are less dramatic than
for the coherence (see \fig{density_r_inset_071}). Within mean-field
theory,  we have compared the density profiles  of the finite system
directly with those in the thermodynamic limit (LDA), using the
finite $N$ solution obtained by the adapted QMC calculation described
in \sect{mean_field_hamiltonian}.  The mean-field analysis indicates
that correlation effects are at the origin of the size-effects of the
full QMC density profiles in \fig{density_r_inset_071}, in particular,
at small system size, $N=1000$.

\subsection{Bose--Einstein cross-over}

The finite-size effects in the coherence reflect the underlying
discrete mode structure of level spacing $\sim  \hbar \omega$.
Off-diagonal properties for $\xi_\Delta \gtrsim l_r$ are cut off by
the extension of the unperturbed ground-state wavefunction, $l_r=
(m\omega/\hbar)^{-1/2}$, and resemble those of a Bose-condensed
system with a significant ground-state occupation.  Whereas in the
thermodynamic limit, the interacting quasi-two-dimensional trapped Bose
gas undergoes a Kosterlitz--Thouless phase transition, the cross-over
to Bose--Einstein condensation sets in when $\Delta  \approx \beta \hbar
\omega$.  If this happens outside the fluctuation regime, $\Delta_{\mf}^f
\lesssim \Delta_{\mf} \lesssim \beta \hbar \omega$, the Bose condensation
will essentially have mean-field character. Since the temperature scale
is given by $\hbar \omega/\TBECZERO =\pi/\sqrt{6N}$,
the discrete level spacing is important for small system sizes,
$N \lesssim N_{\fs}$, with
\begin{equation}
N_{\fs}(\Delta_{\mf}) \approx \frac{1}{6}  \frac{\pi^2 }{
\Delta_{\mf}^2 t^2} = \frac{\pi^2 }{6 \gtilde^2 t^2 (\Delta_{\mf}/\gtilde)^2}.
\elabel{finite_size_N}
\end{equation}
For small $\gtilde$, close to $\TBECZERO$ where $\Delta_{\mf}$ is
of order $\gtilde$, these finite-size effects trigger Bose--Einstein
condensation for small $N$. In particular, for systems with $N\lesssim
N_{\fs}(\Delta_{\mf}^f )\approx \pi^4/(6 \gtilde^2)$, a cross-over
to a mean-field-like Bose condensation occurs\footnote{In contrast
to the infinite mean-field gas, these finite mean-field systems
undergo a Bose--Einstein condensation slightly below $\TBECZERO$
\cite{TKTB}.}, whereas for $N \gtrsim N_{\fs}(\Delta_{\mf}^c)\approx
400 \gtilde^{-2}$ Kosterlitz--Thouless-like behavior sets in (see inset
of \fig{pairs_corr}).  We notice that the finite-size scale $N_{\fs}
\propto 1/g^2$ diverges very rapidly with vanishing interactions, which
could make the cross-over experimentally observable.

For a finite system with $N \lesssim  N_{\fs}(\Delta_{\mf}^f)$, the
condensate wavefunction does not develop immediately a Thomas--Fermi
shape, but remains close to the Gaussian ground-state wavefunction of
the ideal gas with typical extension $l_r=(m \omega/\hbar)^{-1/2}$.
Thus, for small condensate fraction $n_0$, deviations of the moment of
inertia $I$ of the trapped gas from its classical value, $I_{\cl}=\int
\text{d}^2{\rvec} \, r^2 n(r) \sim N l_T^2$, are negligible, of order
$(I_{\cl}-I)/I_{\cl} \sim n_0 l_r^2/l_T^2 \sim N^{-1/2}$.  Only for
larger condensates with $\gtilde N_0 \gg 2 \pi$, the self-interaction
energy dominates the kinetic energy, and the condensate wavefunction
approaches the Thomas--Fermi distribution of  radius $ \sim l_T$,
resulting in  a non-classical value of the moment of inertia.  In this
low-temperature regime, the system can be described by a condensate with
a temperature-dependent fluctuating phase \cite{dima2000}.  Therefore,
for small systems, a non-classical moment of inertia only occurs  at
lower temperatures than condensation, roughly, at a condensate fraction
$n_0 \gtrsim \gtilde$.

To illustrate the cross-over between the Bose--Einstein regime at small
$N$ and the Kosterlitz--Thouless regime at large $N$, we have calculated
the condensate fraction and condensate wavefunction for the ENS parameter
in \fig{n0_ENS}.  To determine both quantities in inhomogeneous systems,
$n^{(1)}_{3d}(\rvecD,\rvecD')$ must be explicitly diagonalized,
as the eigenfunctions of the single-particle density matrix are not
fixed by symmetry alone.  In quasi-two-dimensional systems, the full
resolution of the off-diagonal density matrix in the tightly confined
$z$-direction is difficult. It is more appropriate to consider the
in-plane density matrix, $n^{(1)}(\rvec,\rvec')$, where the confined
direction is integrated over.  $n^{(1)}(\rvec,\rvec')=\int \dd{z}
\int \dd{z'} n^{(1)}_{3d}(\rvec,z;\rvec',z')$.  Because of rotational
symmetry, $n^{(1)}(\rvec,\rvec')$ is block-diagonal in angular-momentum
Fourier components
\begin{equation}
n^{(1)}(\rvec,\rvec')=\sum_{n=0}^\infty \sum_{l=-\infty}^\infty
N_{nl}
\varphi_{nl}^*(r') \varphi_{nl}(r) e^{l \alpha(\rvec,\rvec')}
\end{equation}
where $\alpha(\rvec,\rvec')$ denotes the angle between $\rvec$ and
$\rvec'$, and $N_{nl}$ is the occupation number of the normalized
eigenmode $\varphi_{nl}$.  The (in-plane) condensate fraction, $n_0
=N_{00}/N$, corresponds to the largest eigenvalue with $l=0$ and
condensate wavefunction $\varphi_0(r)=\varphi_{00}(r)$.  Projection on
the Fourier components is convenient for determining the condensate
fraction and wavefunction within QMC.

For the ENS parameters in \fig{n0_ENS}, the central density is already inside
the fluctuation regime, and the condensate wavefunction differs from
the Gaussian ground state of an ideal gas. However, for small systems,
$N \lesssim 10^3$, it still has a Gaussian shape indicating that the
condensate kinetic energy dominates the potential energy. The condensate
fraction vanishes as $n_0 \sim N^{-1/2}$.

The QMC calculations of \cite{Holzmann_Krauth_07} demonstrated that the
condensate fraction of the quasi-two-dimensional Bose gas vanishes in
the normal and in the superfluid phase in the thermodynamic limit, $N \to
\infty$. However, in the low-temperature superfluid phase, the condensate
fraction approaches zero very slowly with increasing system size, so that
an extensive condensate remains for practically all mesoscopic systems.

\wfigure{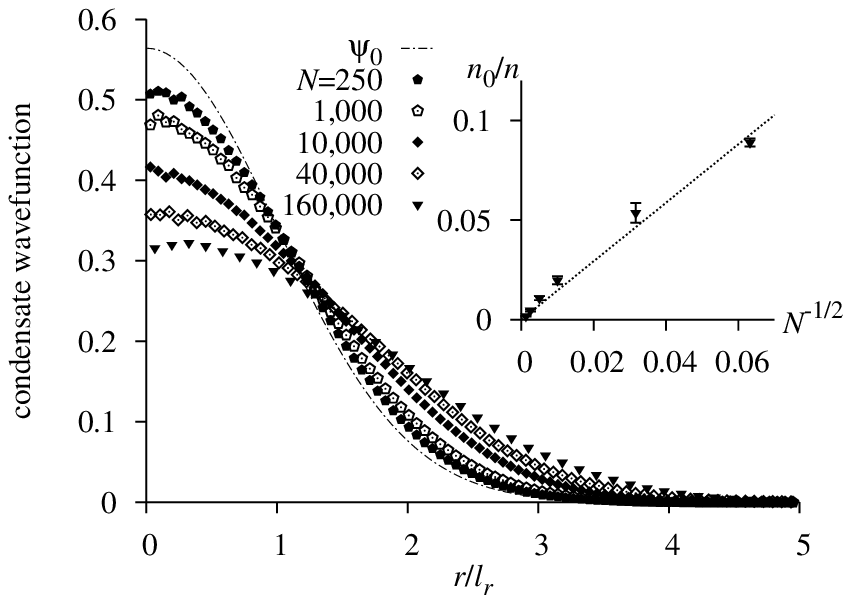}{Condensate fraction $n_0$ (inset) and wavefunction
$\varphi_0(r)$ (main graph), computed by QMC  for ENS parameters with
$t=0.71$ for various system sizes.  The condensate wavefunction is
modified with respect to the ground-state wavefunction $\psi_0(r)$
of the unperturbed harmonic oscillator, but the Gaussian shape is
preserved.  The two largest systems are above the scale $N_\fs$
(see \eq{finite_size_N})}

\section{Two-particle Correlations}
\subsection{Pair-correlation function}
\slabel{2body}

Density--density correlations can be analyzed by considering the
three-dimensional pair-correlation function, $n^{(2)}(\rvecD;\rvecDP)$.
This quantity factorizes within mean-field theory into terms described
by the one-particle density matrix, $n^{(1)}_{\mf}(\rvecD_1;\rvecD_2)$
(see \sect{1body}):
\begin{equation}
n^{(2)}_{\mf}(\rvecD_1;\rvecD_2) = n_{\mf}(\rvecD_1) n_{\mf}(\rvecD_2)
+ \glc n^{(1)}_{\mf}(\rvecD_1;\rvecD_2) \grc^2.
\elabel{pair3Dmf}
\end{equation}
For vanishing distances, $\rvecD_1 \to \rvecD_2$, mean-field
theory predicts $n^{(2)}_{\mf}(\rvecD,\rvecD)=2 n^2_{\mf}(\rvecD)$.
For Bose-condensed atoms, this bunching effect is absent.

In two dimensions, deviations from $2 n^2$ of the
pair-correlation function at contact signal beyond-mean-field
fluctuations\cite{Kagan3,Svistunov2D_2,giorgetti}. In
Ref.~\cite{Svistunov2D_2}, the universal character of the contact value
was used to define the quasi-condensate density, $n_{\qc}(r)\equiv
[2 [n(r)]^2-n^{(2)}(r,r)]^{1/2}$. This quantity has been studied in
Ref.~\cite{blakie} for a quasi-two-dimensional trapped gas within
classical-field theory.

The pair-correlation function of the quasi-two-dimensional gas is obtained
by integrating both coordinates over the confined direction
\begin{equation}
n^{(2)}(\rvec_1;\rvec_2) = \int \dd {z_1} \int 
\dd {z_2} n^{(2)}(\rvec_1,z_1;\rvec_2,z_2), 
\end{equation}
and mean-field expressions for this quantity follow from \eq{pair3Dmf}
together with \eqtwo{nconf}{noff}.  Inside the fluctuation regime,
the gap model (using \eq{gap_model} in the mean-field expressions)
again leads to an improved pair-correlation function $n^{(2)}_\Delta$.

\figg{pair_ENS} illustrates that outside the fluctuation regime mean-field
theory describes the pair-correlation function well.  In contrast to a
strictly two-dimensional gas, the contact value of the pair correlation
function is below $2$.  Even in the mean-field regime, the occupation of
more than one mode in the confined direction causes a noticeable reduction
of the pair-correlation function at contact. The above definition of the
quasi-condensate must therefore be modified in this geometry to maintain
its universal character.

\wfigure{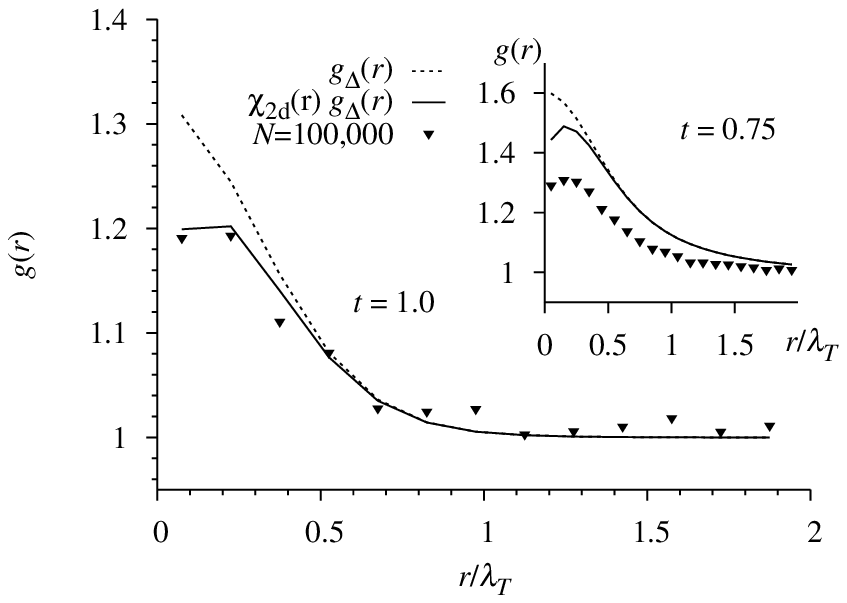}{Central pair correlations,
$g(r)=n^{(2)}(r,0)/[n(r)n(0)]$, of the quasi-two-dimensional
trapped Bose gas at temperature $T=\TBECZERO$ (main figure)
and $T=0.75 \, \TBECZERO$ (inset) for $N=100,000$ atoms (ENS
parameters), together with the prediction of the mean-field gap model,
$g_\Delta(r)=n^{(2)}_\Delta(r,0)/n(r)n(0)$, and the short-range improved
mean-field model, $\chi_{2d}(r) g_\Delta(r)$.}

At short distances $r \sim r_0$, the pair-correlation function depends
on the specific form of the interaction. This cannot be reproduced by
the single-particle mean-field approximation.  However, two-particle
scattering properties dominate for small enough distances as, for example,
the wavefunction of hard spheres must vanish for overlapping particles.
This feature can be included in mean-field theory by multiplying its
pair-correlation functions by a short-range term $\chi_{2d}(r)$, which
accounts for two-particle scattering\cite{pair}. In two dimensions,
$\chi_{2d}$ shows a characteristic logarithmic behavior for short
distances:
\begin{equation*}
\chi_{2d}(r \to 0) \simeq \left[ 1+\frac{\gtilde}{2\pi} \log
\sqrt{\frac{\pi e^C}{2}} \frac{r}{\lD}\right]^2.
\elabel{}
\end{equation*}
Factorizing out the short-range behavior from the pair-correlation
function, the correlation part of the renormalized pair-correlation
function, $\ntilde_\Delta^{(2)}-\ntilde^{(2)}/\chi_{2d}$, should be
dominated by contributions from classical-field theory, its contact
value is universal, and it might be used to define a quasi-condensate
density in quasi-two dimensions via $\ntilde_{\qc}^2= \lim_{r \to
0} \glc \ntilde^{(2)}_\Delta(r,0)-\ntilde^{(2)}(r,0)/\chi_{2d}(r)
\grc$ (see \fig{pairs_corr}).  Similar to the correlation density,
the quasi-condensate density is universal. At $T_\KT$, the classical
field result is $\ntilde_{\qc} \simeq 7.2$, whereas it is around $2.7$
at the onset of the fluctuation regime, so that, in the normal phase,
$n_{\qc}/n$ vanishes as $|\log \gtilde|^{-1}$ for $\gtilde \to 0$.

\wfigure{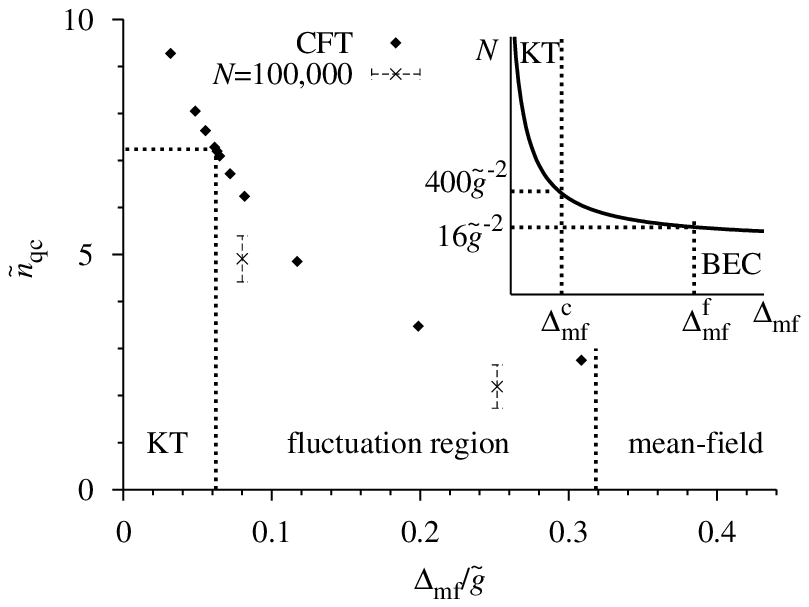}{Quasi-condensate density $n_{\qc}$ obtained by QMC
from the renormalized pair-correlation function at $T=0.71 \, \TBECZERO$,
and $T=0.75 \, \TBECZERO$ (see \fig{pair_ENS}) for ENS parameters,
plotted as a function of $\Delta_{\mf}$, and compared to classical-field
simulations \cite{Svistunov2D_2}. The inset shows the boundary of
the region with strong finite-size effects (see \eq{finite_size_N}). }

\subsection{Local-density correlator}
\slabel{correlator}

Due to the three-dimensional nature of the underlying interaction
potential, observables which couple directly to local three-dimensional
density fluctuations involve the following density correlator
\begin{equation}
K^{(2)}= \lim_{\delta \to 0} \frac{ \sqrt{2 \pi}  l_z \int 
\dd z n^{(2)}(\rvec,z;\rvec+\delta,z)}{\chi_{3d}(\delta)}, 
\end{equation}
where  $\chi_{3d}(r \ll \lD)  \simeq (1-a_s/r)^2$ describes the universal
short-distance behavior of the three-dimensional two-body wavefunction
in terms of the $s$-wave scattering length $a_s$.

Arguments similar to those in \sect{2body} show that the
local-density correlator in general differs from the contact value
of the quasi-two-dimensional pair-correlation function, Further, the
integration over the square of the ground state density in $z$ leads to
a phase-space-density dependence which destroys the  simple mean-field
property $K^{(2)} \propto 2 n$ of strictly two-dimensional Bose gases.

\section{Conclusions}

In this paper, we have studied the quasi-two-dimensional trapped Bose gas
in the normal phase above the Kosterlitz--Thouless temperature for small
interactions $\gtilde <1$.  We have discussed the three qualitatively
distinct regimes of this gas: For phase-space densities $\ntilde
\lesssim 1$, it is classical. At higher density, $1 \lesssim \ntilde
\lesssim \ntilde_f$, the gas is in the quantum mean-field regime, and
its coherence can be maintained over distances much larger than $\lD$.
Finally, mean-field theory fails in the fluctuation regime $\ntilde_f
\lesssim \ntilde \le \ntilde_c$ and beyond-mean-field corrections must
be taken into account.

In the fluctuation regime, for  small interactions, the deviations
of the  density profile with respect to the mean-field profile are
universal. Mean-field theory thus accounts for most microscopic details
of the gas (which depend on the interactions and on the trap geometry).
We have shown in detail how to extract the correlation density (the
difference between the density and the mean-field density at equal
chemical potential) from QMC density profiles and the LDA mean-field
results, and compared it to the universal classical field results.
Quantum corrections to the equation of state, expected of order $\gtilde$,
were demonstrated to be small for current experiments with $\gtilde
\lesssim 0.2$. The smooth behavior of quantum corrections, which has
been already noticed in QMC calculations of the Kosterlitz--Thouless
transition temperature in homogeneous systems \cite{stefano}, strongly
differs from the three-dimensional case\cite{club3/2,n0} where quantum
corrections to universality are non-analytic \cite{NLO,Arnold2}, and
where the universal description holds only asymptotically. It would be
interesting if these universal deviations from mean-field theory could
be observed experimentally.

Correlation effects in local observables, e.g. the density profile, and
local-density correlators, converge rather quickly to their thermodynamic
limit value, and correlation effects for mesoscopic systems are well
described by the local-density approximation.  Off-diagonal coherence
properties show much larger finite-size effects, in particular for
weak interactions.  This introduces qualitative changes for mesoscopic
system sizes and, in particular, the cross-over between Bose--Einstein
physics at small particle number $N \lesssim \const /\gtilde^2$
and the Kosterlitz--Thouless physics for larger systems.  
Tuning the interaction strength via a Feshbach resonance might make it
possible to observe the cross-over between Bose--Einstein
condensation and Kosterlitz--Thouless physics in current experiments.

\acknowledgments
W. K. acknowledges the hospitality of Aspen Center for Physics, where
part of this work was performed.\\

\noindent
NB: The computer programs used in this work are available from the
authors.


\begin{thebibliography}{99}

\bibitem{Dalibard_2006} Z.~Hadzibabic, P.~Kr\"{u}ger, M.~Cheneau,
B.~Battelier, and J.~Dalibard, {\it Nature} {\bf 441}, 1118 (2006).

\bibitem{Clade} P. Clade, C.  Ryu, A. Ramanathan, K. Helmerson, W.D. Phillips
{\it Phys. Rev. Lett.} {\bf 102}, 170401 (2009).

\bibitem{KT} J.~M.~Kosterlitz and D.~J.~Thouless, {\it J. Phys. C}
{\bf 6}, 1181 (1973); J.~M.~Kosterlitz, {\it J. Phys. C} {\bf 7}, 1046
(1974); V.~L.~Berezinskii, {\it Sov. Phys. JETP} {\bf 32}, 493 (1971);
{\bf 34}, 610 (1972).

\bibitem{Dalibard_2007} P.~Kr\"{u}ger, Z.~Hadzibabic, and J.~Dalibard,
{\it Phys. Rev. Lett.} {\bf 99}, 040402 (2007).

\bibitem{Holzmann_Krauth_07} M.~Holzmann and W.~Krauth, {\it
Phys. Rev. Lett.} {\bf 100} 190402 (2008).

\bibitem{HCK} M.~Holzmann, M. Chevallier, and W.~Krauth, {\it
EPL} {\bf 82} 30001 (2008).

\bibitem{Hohenberg} D.S.~Fisher and P.C.~Hohenberg, {\it Phys. Rev. B} {\bf 37}, 4936 (1988).

\bibitem{TKTB} M.~Holzmann, G.~Baym, J.-P.~Blaizot, and F.~Lalo\"e,
{\it Proc. Nat. Acad. Sci.} {\bf 104}, 1476 (2007).

\bibitem{Svistunov2D} N.~Prokof'ev, O.~Ruebenacker, and B.~Svistunov,
{\it Phys. Rev. Lett.} {\bf 87}, 270402 (2001).



\bibitem{Dali} Z.~Hadzibabic, P.~Kr\"{u}ger, M.~Cheneau, S.~P.~Rath,
and J.~Dalibard, {\it New J. Phys} {\bf 10} 045006 (2008).

\bibitem{BBB} R.~N. Bisset, D. Baillie, and P.~B. Blakie, {\it Phys. Rev. A}
{\bf{79}}, 013602 (2009).


\bibitem{Svistunov2D_2} N.~Prokof'ev and B.~Svistunov, {\it Phys. Rev. A}
{\bf 66}, 043608 (2002).


\bibitem{dima2000} D.~S.~Petrov, M.~Holzmann, and G.~V.~Shlyapnikov,
{\it Phys. Rev. Lett.} {\bf 84}, 2551 (2000).

\bibitem{dima2001} D.~S.~Petrov and G.~V.~Shlyapnikov, 
{\it Phys. Rev. A} {\bf 64}, 012706 (2001).

\bibitem{stoof2008} L.-K. Lim, C. M. Smith, and H.T.C. Stoof, 
{\it Phys. Rev. A} {\bf 78}, 013634 (2008).

\bibitem{qmc} W.~Krauth, {\it Phys. Rev. Lett.} {\bf 77}, 3695 (1996).

\bibitem{mchf} M.~Holzmann, W.~Krauth, and M.~Naraschewski, {\it
Phys. Rev. A} {\bf 59}, 2956 (1999).

\bibitem{SMAC} W.~Krauth, {\it Statistical Mechanics: Algorithms and
Computations}, Oxford University Press (Oxford, UK) (2006).

\bibitem{Kagan3} Yu. Kagan, V.A. Kashurnikov, A.V. Krasavin, N.V. 
Prokof'ev, and B.V. Svistunov, {\it Phys. Rev. A} {\bf 61}, 43608 
(2000).

\bibitem{Chevallier} M. Chevallier and W. Krauth, {\it Phys. Rev. E}
{\bf 76} 051109 (2007).

\bibitem{giorgetti} L. Giorgetti, I. Carusotto, and Y. Castin, {\it
Phys. Rev. A} {\bf 76}, 013613 (2007).

\bibitem{blakie} R.N. Bisset, M.J. Davis, T.P. Simula, and P.B. Blakie, {\it Phys. 
Rev.} A {\bf 79}, 033626 (2009).

\bibitem{pair} M.~Holzmann and Y.~Castin,  {\it Eur. Phys. J. D}  {\bf
7}, 425 (1999).

\bibitem{stefano} S. Pilati, S. Giorgini, and N. Prokof'ev, {\it
Phys. Rev. Lett.} {\bf 100}, 140405 (2008).

\bibitem{club3/2} G.~Baym, J.-P.~Blaizot, M.~Holzmann, F.~Lalo\"e, and
D.~Vautherin, {\it Phys. Rev. Lett.} 83, 1703 (1999); {\it Eur. Phys. J.}
B24, 107 (2001).

\bibitem{n0} M.~Holzmann and G.~Baym, {\it Phys. Rev. Lett.} 90, 040402
(2003).

\bibitem{NLO} M.~Holzmann, G.~Baym, J.-P.~Blaizot, and F.~Lalo\"e,
{\it Phys. Rev. Lett.} 87, 120403 (2001).

\bibitem{Arnold2} P. Arnold, G. Moore, and B. Tom\'as\^ik, {\it
Phys. Rev. A} {\bf 65}, 013606 (2002).

\end{thebibliography}
\end{document}